\documentclass[preprint,showpacs,showkeys,aps,prb]{revtex4}
\usepackage[dvips]{graphicx}

\begin{document}

\title{Nonequilibrium Fluctuations in a Gaussian Galton Board \\
(or Periodic Lorentz Gas) Using Long Periodic Orbits}
\author{Wm. G. Hoover and Carol G. Hoover                \\
Ruby Valley Research Institute                           \\
Highway Contract 60, Box 601                             \\
Ruby Valley, Nevada 89833                                \\
}

\date{\today}



\begin{abstract}

Predicting nonequilibrium fluctuations requires a knowledge of nonequilibrium
distribution functions.  Despite the distributions' fractal character some
theoretical results, ``Fluctuation Theorems'', reminiscent of but distinct
from, Gibbs' equilibrium statistical mechanics and the Central Limit Theorem,
have been established away from equilibrium and applied to simple models.  We
summarize the simplest of these results for a Gaussian-thermostated Galton
Board problem, a field-driven mass point moving through a periodic array of
hard-disk scatterers.  The billion-collision trillion-timestep data we analyze
correspond to periodic orbits with up to 793,951,594 collisions and
447,064,397,614 timesteps.

\end{abstract}

\maketitle

\begin{figure}
\vspace{1 cm}
\includegraphics[height=4cm,width=5cm,angle=-90]{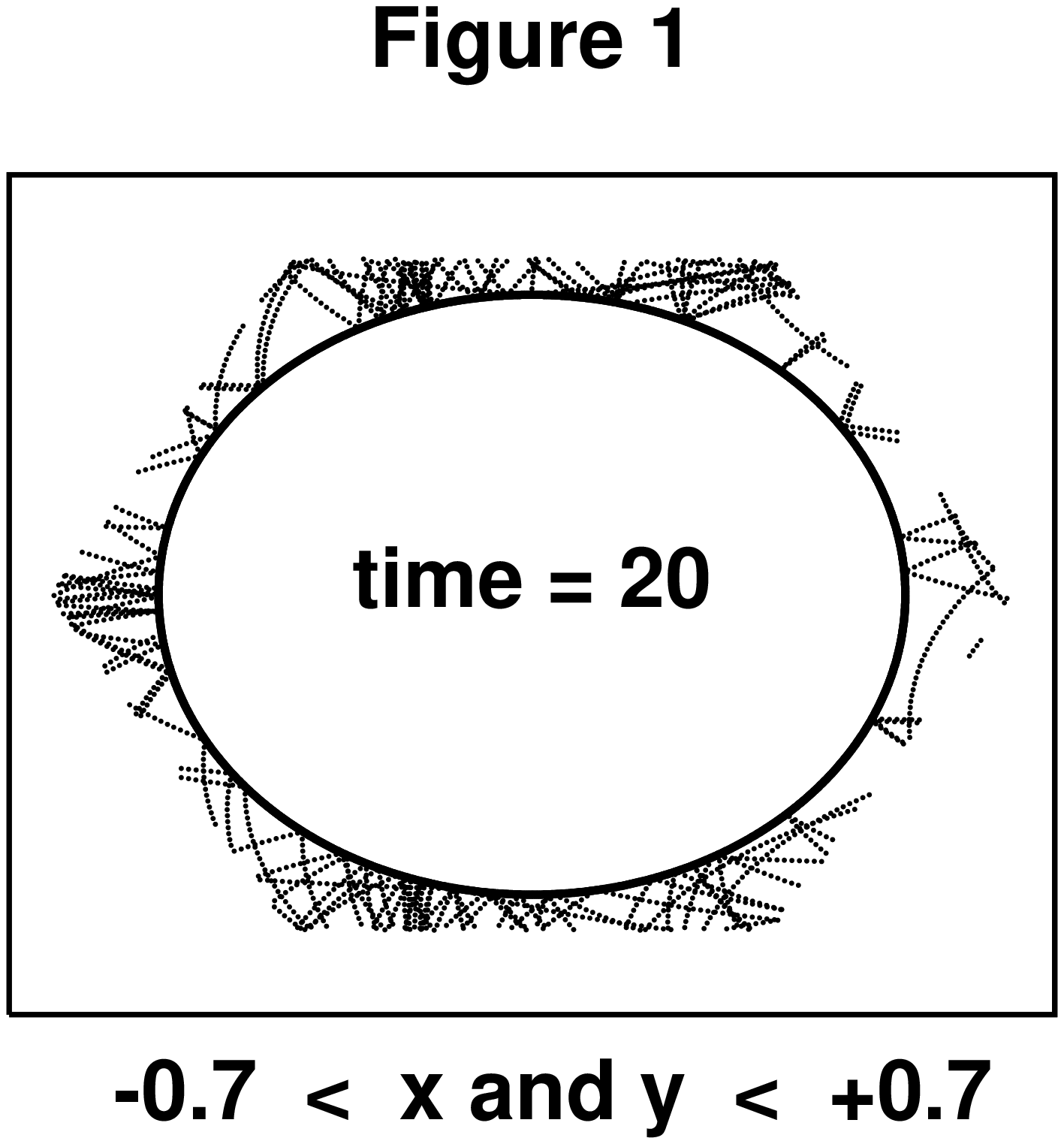}
\caption{
A periodic hexagonal unit cell description of the thermostated Galton Board.
A point particle is scattered by the disk of unit diameter at the cell center.
The scatterer density is 4/5 the close-packed density, so that the
center-to-center spacing of the scatterers is $\sqrt{5/4}$. The accelerating
field, $E = 3$, is directed toward the right, in the horizontal $x$ direction.
The preponderance of collisions at the lefthandside of the scatterer reflects
the resulting positive current, which has a mean value $\langle p_x \rangle =
0.220$.   The magnitude of the velocity is unity so that the instantaneous
current always lies between $-1$ and $+1$.  Accordingly, the time-averaged
entropy production rate is
$\langle \sigma \rangle /k = \langle \dot S \rangle /k = E \langle p_x \rangle /kT = 0.660$.
The combined
length of the trajectory segments shown in the Figure, 20, is equal to the
elapsed time.  A coarse-grained (twelve-digit) description of the model with
a fourth-order Runge-Kutta timestep $dt = 0.0005$ results in a
 793,951,594-collision periodic-orbit problem as is discussed in the text.
}
\end{figure}

\section{Introduction}

In preparing a second edition of {\em Time Reversibility, Computer Simulation,
and Chaos}\cite{b1} we are presently summarizing some of
the recent work in this field in a pedagogical form.  We would appreciate
readers' suggestions as to topics which ought to be included or expanded.
One such topic is considered here, ``Fluctuation Theorems''.

By now there is a voluminous literature devoted to Fluctuation
Theorems of the type described first in 1993 by Evans, Cohen, and
Morriss\cite{b2,b3}.  These theorems relate the relative probabilities of
sufficiently-long forward and reversed nonequilibrium trajectory segments
to the corresponding external entropy produced along the forward
trajectory\cite{b2,b3,b4,b5,b6}.  The time-reversibility of deterministic
thermostated motion equations simplifies such calculations.

Among the simplest applications is the ``Galton Board'' problem, the
field-driven motion of a point particle through a periodic array of hard-disk
scatterers\cite{b7,b8}.  We illustrate that application here as a
worked-out pedagogical exercise problem.  This problem makes contact with
other areas of the research literature: periodic orbit
analysis\cite{b9,b10,b11,b12,b13} and the effects of finite precision on
simulation results\cite{b9,b10}.  We simplify the analysis by considering a
phase-space distribution representing a single periodic orbit.  The orbit
is long enough (millions of collisions and billions of timesteps) to
closely approximate a full nonequilibrium ensemble average.  The orbit
lengths used appear in Table I.  They are sensitive to the exact 
details of the trajectory calculation.  Related examples of the underlying
Galton Board problem have been discussed at length in the
literature\cite{b4,b5,b6,b13,b14}.

\vspace{2 cm}

Table I.  Number of decimal digits $n$, number of collisions, number of
timesteps, and collision rate $\Gamma$  in typical periodic orbits
where each collision is centered in a phase-space cell described with a
spacing of $n$ decimal digits.  The fourth-order Runge-Kutta timestep is
0.0005.  The correlation dimension $D_2 = 1.583$ from Reference 19
predicts orbit lengths of order $10^{0.79n} \simeq 3\times 10^9$ for $n=12$.
\begin{table}[th] 
\begin{tabular}{|c|c|c|c|}
\hline
$n$ & collisions & timesteps & $\Gamma$      \\   
\hline
3  & 774         & 440 812         & 3.512 \\ 
4  & 10 175      & 5 794 556       & 3.512 \\ 
5  & 5 133       & 2 886 067       & 3.557 \\ 
6  & 53 042      & 29 911 691      & 3.547 \\ 
7  & 77 418      & 43 668 154      & 3.546 \\ 
8  & 5 004 959   & 2 819 006 271   & 3.551 \\ 
9  & 2 946 042   & 1 660 446 602   & 3.548 \\ 
10 & 18 398 545  & 10 359 262 120  & 3.552 \\ 
11 & 85 030 972  & 47 885 512 832  & 3.551 \\ 
12 & 793 951 594 & 447 064 397 614 & 3.552 \\ 
\hline
\end{tabular}  
\end{table}

\begin{figure}
\vspace{1 cm}
\includegraphics[height=8cm,width=5cm,angle=-90]{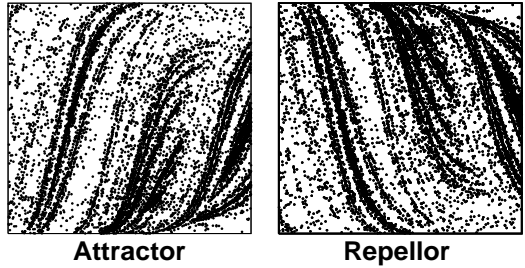}
\caption{
10 175-point periodic attractor and the corresponding (mirror-image)
repellor, using four decimal digits to divide the collision space shown
into $10^4 \times 10^4 = 10^8$ separate states.  The abscissa is
$0 < \alpha < \pi $ and the
ordinate is $-1 < \sin (\beta) < +1$;  $\alpha$ gives the location of
a collision relative to the field direction, while $\beta$ gives the
angle between the post-collision velocity vector and the outward
normal vector at the collision location.  Reversing the time corresponds
to changing the sign of the ordinate.
}
\end{figure}

\section{Background}
Consider the longtime phase-space probability density $f(q,p,\zeta)$
generated by  motion in a nonequilibrium steady state.  Besides the details
of the time-dependent coordinates $\{q\}$ and momenta $\{p\}$, included also
is at least one thermostat variable $\zeta$, which defines the external
time-dependent entropy production rate $\sigma = \dot S$ required to maintain
the steady
state. When Nos\'e-Hoover thermostats are used the friction coefficient(s)
$\{ \zeta \}$ are independent variables, obeying their own differential
equations:
$$
\{ \ F_{\rm NH} \equiv  - \zeta p \ \} \ ; \
\{ \ \dot \zeta \propto \sum[(p^2/mkT) - 1] \ \} \ .
$$
and guaranteeing that the thermostated momenta included in the sum(s) have
longtime average kinetic temperature(s)  $\{ T \}$.

In the one-body Galton Board simulation problem analyzed in Sections
III and IV $\zeta$ is not an independent variable, but is instead an explicit
linear function of the momentum, a ``Gaussian thermostat'', keeping the
kinetic temperature constant:
$$
 F_{\rm G} \equiv  - \zeta p \ ; \
\zeta = \zeta(p) \propto p_x \
 \longrightarrow \ p^2 \equiv mkT .
$$

To simplify our analysis we consider here computational models ``solved''
by generating finite-difference approximations to their system trajectories.
Finite-difference schemes in bounded phase spaces eventually begin to
repeat their history.  Estimating the number of steps both prior to and
during the repetition is analogous to solving the ``Birthday Problem'', ``How
large must a randomly-chosen group of people be to make it likely that two
share the same birthday?''.
Similarly, random jumps in an $N$-state phase space suggest a longest
periodic orbit length of order $\sqrt{N}$ jumps, a second-longest orbit
shorter by a factor of $e$, a third-longest orbit shorter by $e^2$, and so
on.  Thus in practice there are only a few ($\simeq \ln \sqrt{N}$)
{\em periodic orbits} in a
finite-precision phase space\cite{b9,b10} with their combined lengths less
than twice that of the longest periodic orbit.  In a {\em nonequilibrium}
situation, with a multifractal attractor having a reduced phase-space
dimensionality, there are even-fewer, even-shorter paths.  The details can
be expressed in terms of the multifractal distribution's ``correlation
dimension'', which gives the dimensionality of nearby pairs of trajectory
points\cite{b9,b10}.

By considering only the longest most-likely of these numerical orbits, the
resulting ``natural measure'' in the space is a constant, $f = 1/\Omega $
at each of the $\Omega$ discrete points of the orbit and is zero elsewhere
in the space.  In a typical nonequilibrium steady state the length of this
longest single orbit exceeds the combined lengths of all the rest in the
fine-mesh limit\cite{b9,b10}.  With double-precision arithmetic the typical
mesh size is of order $10^{-14}$.

All of the longest-orbit system variables, including $\sigma $ the rate of
external entropy production due to the thermostat, are necessarily periodic
functions of time with period $\tau$, the orbit length.  The external
entropy produced per period is a positive constant,
$\tau \langle \sigma \rangle $.  Because the nonequilibrium motion equations
underlying our continuous-time problem are all {\em time-reversible}, we can
also usefully imagine a highly-improbable time-reversed {\em backward}
version of the periodic orbit.  See Figure 2 for a four-digit example.  This
``repellor'' trajectory, the time-reversed attractor, is a bit artificial.
It can be generated in either of two  fully-equivalent ways: [1] solve the
differential equations for $\{q,p,\zeta\}$ as usual, but with a {\em negative
timestep} $dt \rightarrow - dt$ or [2] take the stored solution of
the equations with a positive $dt$ and change the signs of the $\{p\}$
and $\{\zeta\}$.  Because a finite bounded phase-space distribution requires
that the Lyapunov instability\cite{b12,b13} of the reversed orbit necessarily
exceeds that of the forward attractor, the reversed orbit can only be
generated in the two ways just mentioned.

Fluctuation Theorems describe the {\em relative probability} of
finite-but-large {\it segments} of such forward-backward pairs.  Unlike the
Central Limit Theorem, which predicts the longtime (Gaussian) {\em shape} of
the probability distribution, the Fluctuation Theorems instead predict ratios
of forward/backward probabilities.  By
considering the simplest computational case where the phase-space motion is
periodic, but dissipative, the discussion of this single-orbit problem
avoids the need to address ergodicity as well as sign changes in the values
of the local (coordinate-dependent) Lyapunov exponents.

Consider an observation time $\delta \tau$ (perhaps as small as a single
timestep and possibly as large as the total length $\tau $ of the periodic
orbit under consideration).  Averaging the location of the observation time
over the entire orbit gives exactly the same rate of external entropy
production (due to the thermostat) as characterizes the full orbit: 
$$
\langle \sigma \rangle_{\delta \tau} =
(1/\delta \tau)\Delta S_{\delta \tau} \equiv (1/\tau)\Delta S_\tau =
\langle \sigma \rangle_{\tau} 
= (1/\tau)(work/T)_{\rm orbit} = (1/\tau)(heat/T)_{\rm orbit} \ .
$$
The $work$ done (by a driving external field), summed up over the
entire orbit, is necessarily equal to the total $heat$ extracted by the
constant-temperature thermostat.  Dividing by the thermostat temperature
$T$ gives the corresponding entropy produced,
$\Delta S =(heat/T)_{\rm orbit}$.  In the special
case we consider in Sections III and IV the kinetic temperature is
fixed by using a ``Gaussian'' thermostat.  Gauss' {\em Principle of Least
Constraint} provides a basis for this approach.  The Principle
suggests using the smallest possible rms force to constrain the kinetic
temperature T.
This ``least'' force is linear in the
momentum.  We define the kinetic temperature $T$ in the usual way: $T =
p^2/mk = 1$.  Fluctuation Theorems with fluctuating temperatures and with
stochastic thermostats have also been considered and tested\cite{b4,b5,b6,b15}.

The relative probabilities of the forward ``attractor'' and reversed 
``repellor'' orbits (if we now imagine them as the two infrequently 
communicating parts
of an ergodic steady continuous distribution) can be expressed in terms of
their orbit-averaged Lyapunov exponents.  The entire spectrum of Lyapunov
exponents, both positive and negative, can be determined using a
finite-difference algorithm, as described by Bennetin\cite{b16}, or by
using continuous-time Lagrange multiplier constraints\cite{b17,b18}.  The
positive exponents,which
describe spreading, can be used to express the loss rate of probability
density from the neighborhoods of the forward and backward orbits.  These
loss rates for the attractor $A$ and repellor $R$ must balance in a
steady state.  Averaged over a single periodic orbit, this balance
expresses the attractor and repellor probabilities in terms of the
dissipation induced by the thermostat:
$$
f_A\exp [\sum_{\lambda_A > 0} -\lambda_A\tau] =
f_R\exp [\sum_{\lambda_R > 0} -\lambda_R\tau] 
\longleftrightarrow \frac{f_A}{f_R} = \frac{e^{\sum \lambda_A\tau}}
{e^{-\sum \lambda_R\tau}} = e^{\langle \dot S\rangle \tau/k} \ .
$$
Because the positive exponents on the repellor are simply reversed-sign
versions of the negative exponents on the attractor the two
Lyapunov-exponent sums can be combined:
$$
\ln \biggl[\frac{f_{\rm forward}}{f_{\rm backward}}\biggr] =
\ln \biggl[\frac{f_A}{f_R}\biggr] = 
\sum_{\lambda_A > 0} \lambda_A\tau - \sum_{\lambda_R > 0} \lambda_R\tau
\equiv
\sum_A \lambda \tau = \langle \dot S \rangle\tau /k \ . \ \ \ [FT]
$$
The usual statement of this Fluctuation Theorem [FT] includes the proviso that
the averaging time $\tau $ must be sufficiently large. It is evident that
the steady state quotient $f_A/f_R$ is typically positive, as the Second
Law states, so that the longtime expression $[FT]$ fails as $\tau$
approaches zero.

For Gauss' or Nos\'e-Hoover thermostats the equality between the complete
sum of all the local Lyapunov exponents and the external rate of entropy
production is an identity.  For the
Galton Board example which we detail in Section III this equality follows
directly from an application of Liouville's Theorem to the nonHamiltonian
equations of motion suggested by Gauss' Principle.

The Fluctuation Theorem illustrated here was first demonstrated, numerically,
for a manybody shear flow\cite{b2}.  We illustrate the same Theorem in the
next Section for a simple pedagogical example, the thermostated one-particle
Galton Board\cite{b1,b4,b6,b7,b8,b10,b13,b14,b19,b20}.  We divide up a single
relatively-long finite-precision periodic orbit into portions $\delta \tau$.
Evidently the overall averaged dissipation rate for these portions is the same
as the rate for the entire orbit $\langle \sigma \rangle_\tau $ so that we can
test the applicability of the  Theorem as a function of the sampling time
$\delta \tau$.

\section{Galton Board}

The Galton Board problem provides an instructive example of all these
ideas.  A point particle  with unit mass is accelerated to the right by a
field $E$
through a triangular lattice of fixed disk scatterers.  For this problem
the average current $\langle p_x \rangle$, dissipated energy, and
entropy production are all
simply related:
$$
p^2 \equiv kT \equiv 1 \longleftrightarrow
E\langle p_x \rangle = \langle (d/dt)work \rangle = 
\langle (d/dt)heat \rangle = \langle \zeta p^2 \rangle \ .
$$
The speed $|p/m|$ of the point particle, as well as its ``temperature'' $p^2/mk$,
is kept constant by the friction coefficient $\zeta = Ep_x/p^2$:
$$
\zeta  = (d/dt)work/kT = (d/dt)heat/kT = \dot S/k  \ .
$$
Here $\dot S = \sigma $ is the instantaneous external entropy production rate.
The complete set of motion equations for the isokinetic Galton Board is the
following:
$$
\dot x = p_x \ ; \ \dot y = p_y \ ; \
\dot p_x = F_x + E - \zeta p_x \ ; \ \dot p_y = F_y -\zeta p_y \ .
$$
By switching to polar momentum coordinates these trajectory equations can
be integrated analytically\cite{b7}, though here we choose to use the
equally accurate (machine accuracy) fourth-order Runge-Kutta integration
for simplicity's sake.  The hard-disk elastic force $F$ is the reflective
interaction of the point particle and the fixed scatterer, where the
collision location and direction are given by the angles $\{ \alpha,
\beta \}$ defined in the caption of Figure 2.
The collisional ``jumps'' in the phase-space orbit contribute
to the Lyapunov instability of the problem, but make no contribution to
the work done by the field or to the heat extracted by the thermostat
and converted to external entropy production.  In the numerical work the
coordinates and momenta are rescaled (with $m$, $k$, and the scatterer
diameter all equal to unity),
$$
x^2 + y^2 \longrightarrow 0.25 \ ; \ p_x^2 + p_y^2 \longrightarrow 1 \ ,
$$
whenever the accurate Runge-Kutta trajectory returns $\{x,y\}$ values
inside the scatterer radius of 1/2.

\begin{figure}
\vspace{1 cm}
\includegraphics[height=8cm,width=6cm,angle=-90]{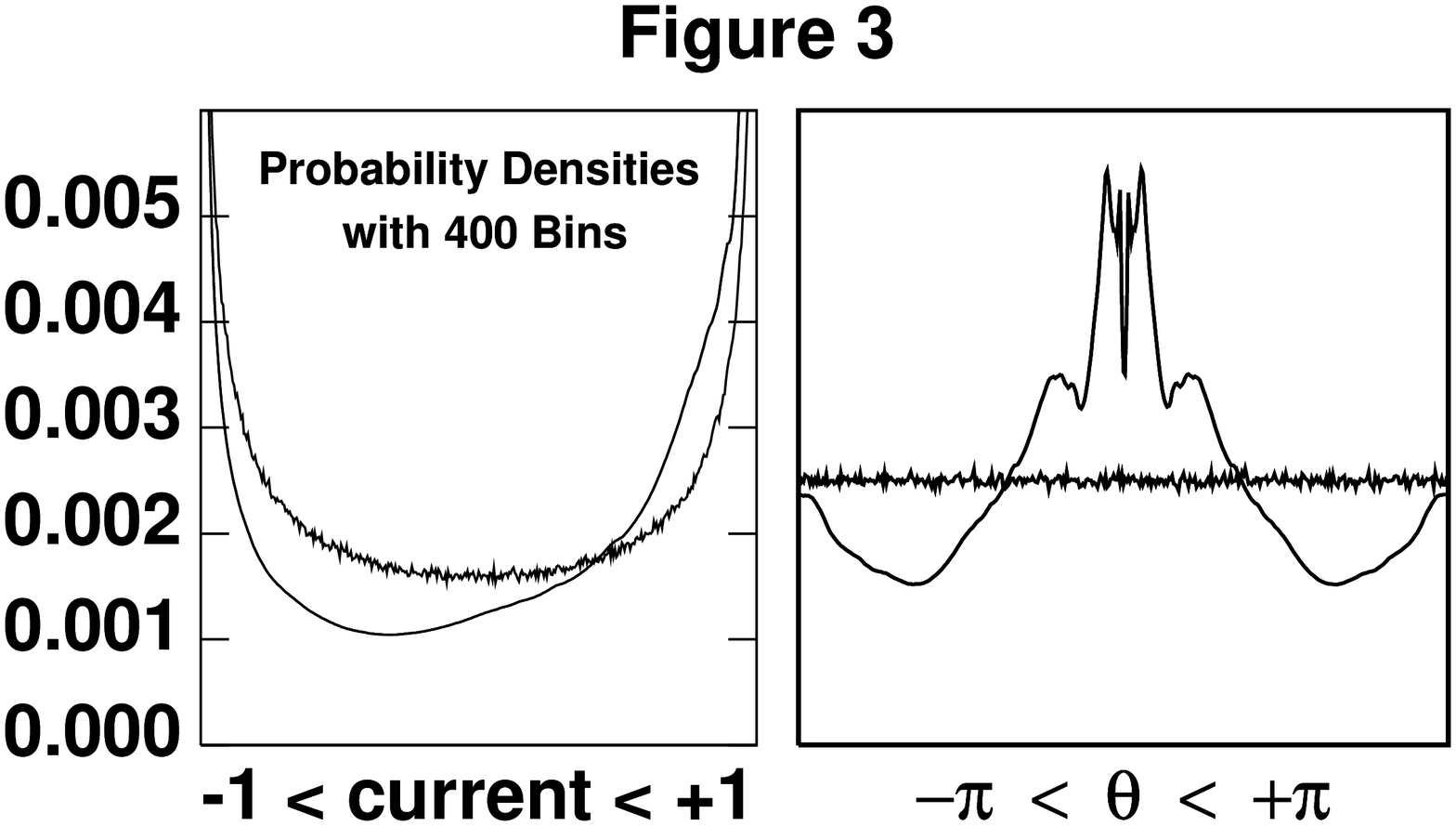}
\caption{
400-bin probability distributions for the current and for the direction of the
velocity, varying from parallel to antiparallel,
$\theta = \arctan (p_y/p_x)$ for no field (jagged symmetric data) and for
a field strength of 3.00.  The relatively complex peak at $\theta = 0$
corresponds to the enhanced
probability on the lefthandside of the scatterer in Figure 1.  A
timestep of $0.0005$ with 100 billion timesteps was used in accumulating
these data.
}
\end{figure}

We apply this model to the Fluctuation Theorem by considering the
situation indicated in Figure 1 for a periodic unit cell.  The Figure
shows an illustrative trajectory portion made up of $20 \ 000$
timesteps, with $dt=0.001$.  In the equilibrium case, with zero field,
the scatterer collisions make all velocity directions equally probable
so that the probability density for $p_x = \cos(\theta)$ diverges at
$\pm 1$:
$$
\frac{d\theta}{2\pi} = {\rm prob}(\theta)d\theta = 
{\rm prob}(p_x)dp_x \rightarrow
$$
$$
{\rm prob}(p_x) = \frac{(|d\theta/dp_x|)}{2\pi} = \frac{1}{2\pi|\sin(\theta)|}
= \frac{1}{2\pi \sqrt{1 - p_x^2}} = \frac{1}{2\pi |p_y|} \ .
$$
With the field turned ``on'' the downhill directions become
more probable, as is illustrated by the trajectory segment of Figure 1
and by the two probability densities, normalized for 400 momentum bins, shown
in  Figure 3.  With the field ``off'', and all velocity directions equally
likely the probability density for $p_x$ diverges at the extrema,
$p_x = \pm1$.

The low-field dynamics is Lyapunov unstable\cite{b20}, with two nonzero
Lyapunov
exponents, $\{ \lambda \} = \{ \pm 3.922\}$ at zero field, and $\{ \lambda \} =
\{ 3.000 \ ;-3.658 \}$ with a field strength of $E = 3.00$.  These
data for the Galton Board, and many others, for simple models and for
manybody systems, are available in Christoph Dellago's 1995
Dissertation\cite{b20}.

If the field strength is large enough, short periodic orbits with both
exponents negative (20 collisions for $E=3.69$ and 2 collisions for $E =
4.00$) can be stabilized in the infinitesimal-mesh limit.  See  Figures
2 and 5 of Reference 7.  To avoid such nonergodic situations we choose a
field strength $E=3.00$, for which the conductivity (current divided by
field) is 0.0734, significantly reduced from the lowfield\cite{b14}
Green-Kubo value of 0.10, and corresponding to a current
$0.0734 E $
$$
\langle p_x \rangle = p\langle \cos (\theta) \rangle =
0.0734 \times 3 = 0.220 \ ,
$$
and a mean squared current of 0.574. These latter numerical results were
obtained in 1987\cite{b7}.

The probability densities for four different sampling times are shown in
Figure 4.  The longest time shown corresponds to approximately 178 collision
times, while the shortest is about 1/6 of a collision time.  Let us turn to
the analysis of the sampling-time dependence of these results from the
longtime standpoints of the Fluctuation Theorem and the Central Limit Theorem.

\section{The Fluctuation and Central Limit Theorems}

The ``Fluctuation Theorem'' expresses the ratios of probabilities of
forward and reversed processes, but not their shapes, ending up with
expressions like this:
$$
\ln\biggl[\frac{{\rm prob}_f(+\sigma)}{{\rm prob}_b(-\sigma)}\biggr]_{\delta \tau} =
\frac{+\sigma \delta \tau}{k} \ ,
$$
valid in the limit that $\delta \tau$ is sufficiently large.  The Central Limit
Theorem, also valid for large $\delta \tau $, can be expressed similarly:
$$
\ln\biggl[\frac{{\rm prob}_f(+\sigma)}{{\rm prob}_b(-\sigma)}\biggr]_{\delta \tau} =
-\frac{(+\sigma - \langle \sigma \rangle)^2}{2\Sigma^2}
+\frac{(-\sigma - \langle \sigma \rangle)^2}{2\Sigma^2} =
+\frac{2\sigma \langle \sigma \rangle}{\Sigma^2} \ ,
$$
where the average entropy production here is $\langle \sigma \rangle = 0.22E =
0.66$ and
$\Sigma $ is the ``standard deviation'' of the Gaussian.  Equating the two
expressions (Fluctuation Theorem and Central Limit Theorem) gives an explicit
large-$\delta \tau$ expression for $\Sigma$:
$$
\Sigma = \sqrt{2k\langle \sigma \rangle/\delta \tau} \ .
$$

  A visual inspection of the current probabilities for a
relatively large time averaging interval $\delta \tau = 50$ (nearly
200 collisions) reveals noticeable deviations from a smooth Gaussian shape.
Much larger intervals are not practical because the probability of observing
negative currents becomes small. For example, for a time interval of $\delta
\tau =100$, where we never observed a ``negative entropy production'' in our
Table I sample length of $2.2\times10^6\delta \tau$,
the probability of the zero-current Gaussian relative to its maximum (at an
entropy production rate of 0.66) is
$$
\exp[-0.66^2/2\Sigma^2] = \exp[-(0.66/4) \times 100]
= \exp[-16.5] \simeq 7\times 10^{-8} \ .
$$

For this example problem it is evident that the two longtime relations are
only semiquantitative (with errors of a few percent) and don't give the
detailed shape of the probability distribution.  To illustrate the Fluctuation
Theorem relationship in the usual way we plot the (logarithm of the)
probability ratio for a range of sampling times, from $2000dt$ to
$10^5dt$.  The data shown in Figures 5 and 6, all for a single typical
12-digit periodic orbit, demonstrate that the Fluctuation Theorem, like the
Central Limit Theorem, is indeed a useful semiquantitative guide provided
that the sampling time is more than a few collisions and that the entropy
production rate is not too large.

\begin{figure}
\vspace{1 cm}
\includegraphics[height=8cm,width=6cm,angle=-90]{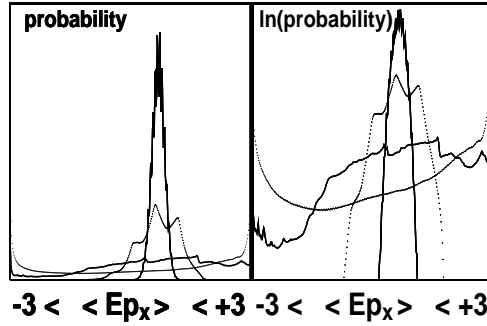}
\caption{
Entropy production rate averaged over averaging time intervals $\delta \tau =
\{50, 5, 0.5,0.05\}$.  The mean time between collisions is 0.282.
}
\end{figure}

\begin{figure}
\vspace{1 cm}
\includegraphics[height=6.4cm,width=8cm,angle=-90]{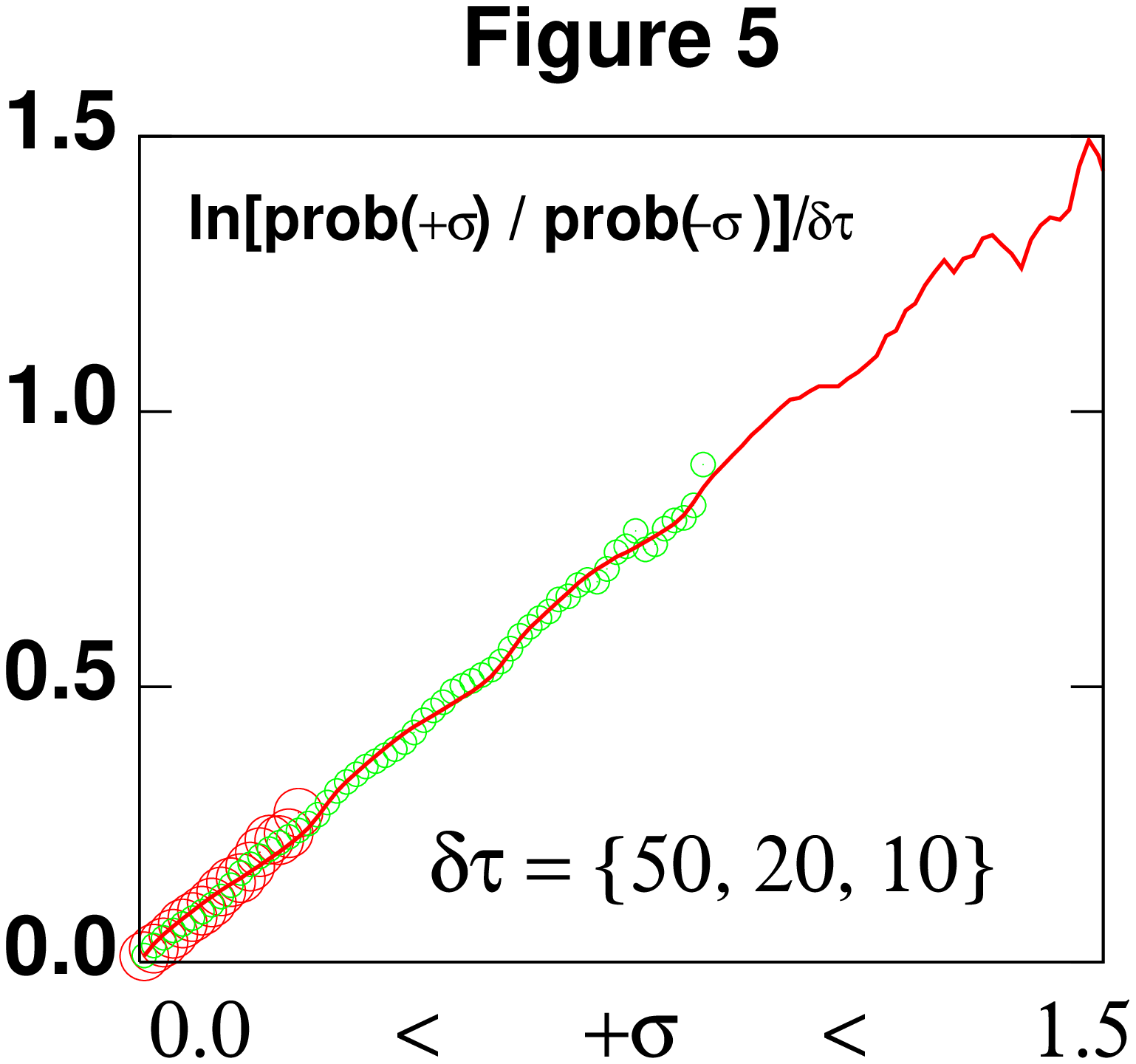}
\caption{
$(1/\delta \tau )\ln [ {\rm prob} (+\sigma/k)/ {\rm prob}(-\sigma/k)]$ as a
function of the entropy production rate $\sigma/k$ averaged over intervals
of length 10 (solid line), 20 (small open circles), and 50 (large open
circles), corresponding to 36, 71, and 178 collision times.  These data
were accumulated from a 12-digit periodic orbit.  According to the
``Fluctuation Theorem'' the slope of this curve is unity for sufficiently
long averaging intervals.  Generating and analyzing these data required just
over a month of machine time.  Boltzmann's constant $k$ is set equal to unity 
in the plot.
}
\end{figure}

\begin{figure}
\vspace{1 cm}
\includegraphics[height=6.4cm,width=8cm,angle=-90]{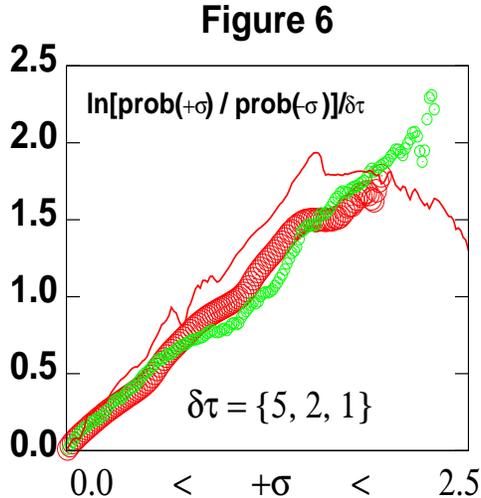}
\caption{
$(1/\delta \tau )\ln [ {\rm prob} (+\sigma/k)/ {\rm prob}(-\sigma/k)]$ as a
function of the entropy production rate $\sigma/k$ averaged over intervals
of length 1 (solid line), 2 (small open circles), and 5 (large open
circles), corresponding to 4, 7, and 18 collision times.  These data
were accumulated from a 12-digit periodic orbit.  According to the
``Fluctuation Theorem'' the slope of this curve is unity for sufficiently
long averaging intervals.  Generating and analyzing these data required just
over a month of machine time.  Boltzmann's constant $k$ is set equal to unity 
in the plot.
}
\end{figure}

\section{Summary}

The Fluctuation Theorem provides accurate estimates for the
relative probability of forward and reversed steady-state
phase-space trajectories.  The Theorem illustrates the usefulness of
coarse-grained probability densities in microscopic interpretations of
macroscopic thermodynamics.  Results for short-term nonequilibrium
fluctuations (most of the data in Figure 4) are highly model dependent, 
and still lack accurate theoretically-based estimates.

The Fluctuation Theorem looks very much like Onsager's (or Gibbs')
relation for probablities in terms of a nonequilibrium phase-space entropy,
$$
{\rm prob} \simeq e^{\Delta S/k} \ ,
$$
even though the nonequilibrium entropy does not exist\cite{b1,b7,b19,b20,b21}
outside the linear-response regime.

The Fluctuation Theorem goes beyond the Central Limit Theorem (which also
applies to nonequilibrium steady states) and so can be used to give an
explicit prediction for the halfwidth of the large-$\delta \tau$ Gaussian
distribution:
$$
\ln \biggl[ \frac{{\rm prob}(+\sigma)}{{\rm prob}(-\sigma)}\biggr]_{FT} =
\delta \tau\sigma /k \ \simeq \
\ln \biggl[\frac{{\rm prob}(+\sigma)}{{\rm prob}(-\sigma)}\biggr]_{CLT} =
2\sigma\langle\sigma\rangle/\Sigma^2 \ ,
$$
where $\Sigma$ is the standard deviation, and accordingly should be
$\sqrt{2k\langle \sigma \rangle/\delta \tau}$.  The two Theorems taken
together do provide a useful semiquantitative guide to nonequilibrium
fluctuations far from the linear-response regime.

The relationship between the length of coarse-grained periodic orbits
and the multifractal correlation dimension can be derived from a
statistical viewpoint, by imagining random jumps among $N$ phase space
states, resulting in an orbit length somewhat less than $\sqrt{N}$.
In the present work the ``jump'' from one collision to the next can
be viewed as such a process.

Many generalizations of this simple isokinetic model have been elaborated
in the literature.  By adding a magnetic field\cite{b4} the time-reversibility of
the equations of motion can be eliminated, but with the results still
obeying the Fluctuation Theorem.  A Nos\'e-Hoover thermostat\cite{b5} allows
for fluctuations in the kinetic energy, but without affecting reversibility.
In both these cases the Fluctuation Theorem is obeyed for sufficiently
large times.  Results in the short-time limit, instantaneous fluctuations
in the entropy production rate, are more highly model dependent and still
cannot be predicted theoretically.

\section{Acknowledgments}

Thomas Gilbert and David Jou kindly provided us with advice and useful
references, including a .pdf copy of Thomas' 15/03/2006 seminar talk,
``Fluctuation Theorem, A Selective Review and Some Recent Results'',
and David's contribution ``Temperature, Entropy, and Second Law
Beyond Local Equilibrium, an Illustration'' to the Proceedings of the
2010 Granada Seminar in La Herradura.  Christoph Dellago provided a copy
of Reference 20 and Denis Evans made some useful comments on the first draft
of this work.  We thank Lakshmi Narayanan at
World Scientific Publishers for her continuing support.

\pagebreak

\end{document}